\documentclass[aps,prl,preprint,amsmath,amssymb,superscriptaddress,nobibnotes,endfloats*]{revtex4-1}
\usepackage{graphicx}
\usepackage{bm}     
\usepackage{hyperref} 
\usepackage{dsfont}    
\usepackage{color}
\usepackage[normalem]{ulem}
\usepackage{cleveref}

\def\A{\mathcal{A}}
\def\B{\mathcal{B}}
\newcommand{\ket}[1]{|#1\rangle}

\newcommand{\bra}[1]{\langle#1|}

\newcommand{\abs}[1]{\left|{#1}\right|}

\newcommand{\eq}[1]{equation~(\ref{#1})}
\newcommand{\fig}[1]{Fig.~\ref{#1}}
\newcommand{\tab}[1]{Table \ref{#1}}

\def\beq{\begin{eqnarray}}
\def\eeq{\end{eqnarray}}

\begin{document}

\title{Observation of detection-dependent multi-photon coherence times }

\author{Young-Sik Ra}
\affiliation{Department of Physics, Pohang University of Science and Technology (POSTECH), Pohang, 790-784, Korea}

\author{Malte C. Tichy}
\affiliation{Physikalisches Institut, Albert-Ludwigs-Universit\"at, Hermann-Herder-Str.~3, D-79104 Freiburg, Germany}
\affiliation{Department of Physics and Astronomy, University of Aarhus, 8000 Aarhus C, Denmark}

\author{Hyang-Tag Lim}
\affiliation{Department of Physics, Pohang University of Science and Technology (POSTECH), Pohang, 790-784, Korea}

\author{Osung Kwon}
\affiliation{Department of Physics, Pohang University of Science and Technology (POSTECH), Pohang, 790-784, Korea}

\author{Florian Mintert}
\affiliation{Physikalisches Institut, Albert-Ludwigs-Universit\"at, Hermann-Herder-Str.~3, D-79104 Freiburg, Germany}
\affiliation{Freiburg Institute for Advanced Studies, Albert-Ludwigs-Universit\"at, Albertstr. 19, D-79104 Freiburg, Germany}

\author{Andreas Buchleitner}
\affiliation{Physikalisches Institut, Albert-Ludwigs-Universit\"at, Hermann-Herder-Str.~3, D-79104 Freiburg, Germany}

\author{Yoon-Ho Kim}
\email[]{Correspondence and requests for materials should be addressed to Y.-H.K. (email: yoonho72@gmail.com).}
\affiliation{Department of Physics, Pohang University of Science and Technology (POSTECH), Pohang, 790-784, Korea}

\date{\today}

\begin{abstract} 
The coherence time constitutes one of the most critical parameters that determines whether or not interference is observed in an experiment. For photons, it is traditionally determined by the effective spectral bandwidth of the photon. Here we report on multi-photon interference experiments in which the multi-photon coherence time, defined by the width of the interference signal, depends on the number of interfering photons and on the measurement scheme chosen to detect the particles. A theoretical analysis reveals that all multi-photon interference with more than two particles features this dependence, which can be attributed to higher-order effects in  the mutual indistinguishability of the particles. As a striking consequence, a single, well-defined many-particle quantum state can exhibit qualitatively different degrees of interference, depending on the chosen observable. Therefore, optimal sensitivity in many-particle quantum interferometry can only be achieved by choosing a suitable detection scheme.
\end{abstract}

\maketitle

The coherence time is the time-scale over which wave-like interference is observed for a particle-like object. In practice, it constitutes the critical parameter that determines whether or not quantum interference is observed in an experiment. It intimately characterizes quantum systems, such as, photons~\cite{PhysRevLett.59.2044,kim03a}, electrons~\cite{Bocquillon}, and molecules~\cite{Hornberger2012}. For photons, the traditional notion of coherence time is  determined by the photon generation process and the bandwidth of spectral filtering~\cite{PhysRevLett.59.2044,keller97}. 

Consider now the  two-photon interference experiment first conducted by Shih--Alley \cite{shih2} and Hong--Ou--Mandel  \cite{PhysRevLett.59.2044} where two single photons impinge on a 50:50 beam splitter (BS)  via different input ports. When the photons are indistinguishable, i.e. when they impinge within their coherence time, they always leave the BS in coalescence, due to the destructive interference of the amplitudes associated with two reflected and two transmitted photons.  If these amplitudes become distinguishable, even  in principle,  interference is lost and the coalescence breaks up \cite{ref:pittman, kim03}. Distinguishability may originate, for instance, from different arrival times \cite{PhysRevLett.59.2044}, polarization states \cite{kwiat92}, group velocity dispersion \cite{PhysRevLett.68.2421}, and spectral distribution \cite{keller97} of the two photons. In such experiments, the  interference signal is observed in coincidence measurements between two single-photon detectors located at the output ports of the BS. When a single-photon detector is placed at each output port and temporal distinguishability is adjusted, the coincidence measurement  exhibits the typical Hong--Ou--Mandel dip \cite{PhysRevLett.59.2044}. Consequently, for two single-photon detectors that are combined with a BS and thus function as a two-photon detector at one output port, 
 a peak appears \cite{kim03a}. The peak and the dip in these two-photon signals have the same width and are  directly related to the single-photon coherence time \cite{PhysRevLett.59.2044,kim03a} that is determined by the spectral properties of the photon~\cite{PhysRevLett.59.2044,keller97}.
  At this stage, the \emph{single}-photon coherence time  seems to adequately characterize the shape of measurable \emph{multi}-photon signals. 

By injecting a two-photon Fock-state \cite{PhysRevLett.83.959} or a three-photon Fock-state \cite{Niu:09} into each input mode of the aforementioned setup and by detecting all photons (four  or six, respectively) at one output port of the BS, similar coincidence peaks have been observed. Defining the multi-photon coherence time as the width of the multi-photon signal, how do  general multi-photon coherence times depend on the single-photon coherence time? More importantly, is the multi-photon coherence time always unambiguously determined by the single-photon coherence time?

 Here we answer these questions experimentally and theoretically. We show in two- and four-photon interference experiments that the measured coherence time depends on the number of interfering photons and on the measurement scheme used to detect the particles.
The unique two-photon coherence time that is determined by the single-photon coherence time turns out to be a rather accidental exception. In particular, an asymmetric four-photon detection scheme leads to a significant reduction of the observed coherence time. A theoretical study reveals that multi-photon signals depend on various powers of the indistinguishability (to be quantified in \eq{eq:indistinguishability} below) of the interfering photons. Since these different terms kick in as soon as more than two photons are brought to interference, our results imply that generic multi-photon  coherence times depend on the number of interfering photons and on the chosen  detection scheme, and, in general, that they are not simply determined by the single-photon coherence time.

\section{Results}

\begin{figure*}[t]
\centerline{\includegraphics[width=85mm]{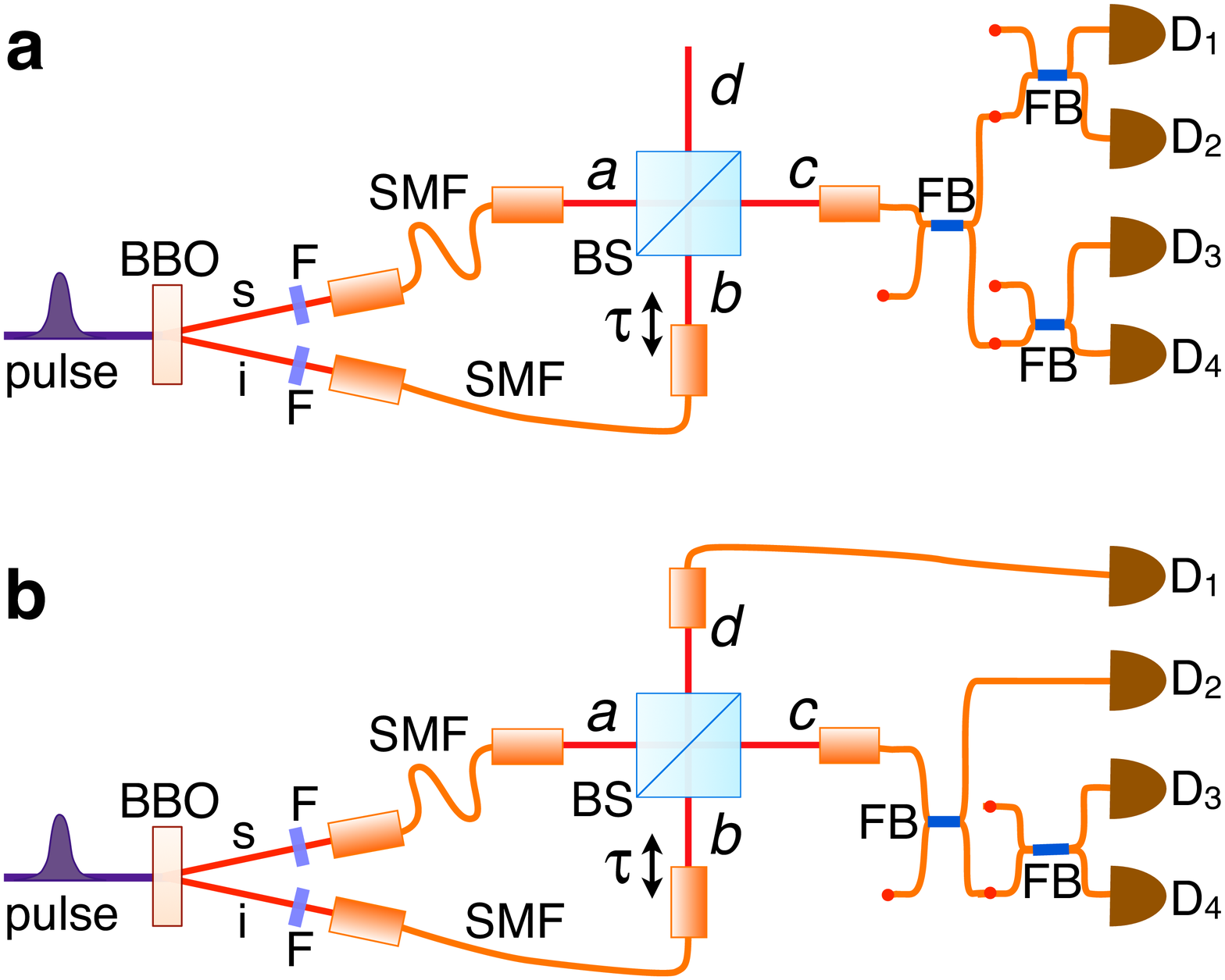}}
\caption{\textbf{Experimental setup.}
A frequency-doubled vertically polarized laser pulse with pulse width of approximately 200 fs (central wavelength: 390 nm, average power: 120 mW) from a mode-locked Ti:Sa laser  is focused onto a type-I $\beta$-barium-borate (BBO, thickness: 2 mm) crystal by a lens (not shown) with a focal length of 300 mm. 
Horizontally polarized signal (s) and idler (i) photons are generated via  non-collinear frequency-degenerate SPDC and filtered by a set of narrow bandpass filters (F). They are centered at 780 nm with FWHM of 4 nm. The signal and idler photons are then coupled into single-mode fibers (SMF) for cleaning up the spatial modes.  The outputs of the SMFs are collimated and compensated for polarization rotation, induced by the propagation through the SMFs, with a set of half-wave and quarter-wave plates (not shown). The photons are then combined at a 50:50 beam splitter (BS). The temporal overlap of the photons at the BS is controlled by adjusting the arrival time delay $\tau$ between the signal and the idler photons with a fiber collimator attached to a motorized stage located at the input mode $b$ of the BS. Photons are then detected at the output modes ($c$ and $d$) of the BS with single-photon detectors D$_1$ $\sim$ D$_4$ (Perkin-Elmer SPCM-AQ4C) coupled to single-mode fibers and 50:50 fiber beam splitters (FB), the coincidence window is less than 8 ns. 
Different detector arrangements permit to detect $m$ and $n$ photons at the output modes $c$ and $d$, respectively, denoted by $(m,n)$-detection. (2,0)- and (4,0)-detection can be observed in \textbf{a} and (1,1)- and (3,1)-detection can be observed in \textbf{b}.
}\label{fig:scheme}
\end{figure*}

{\bf Experimental setup.} The experimental setup for observing such multi-photon coherence times is shown in \fig{fig:scheme}. 
Photon pairs are generated via spontaneous parametric down-conversion (SPDC), they are subsequently led to the input modes $a$ and $b$ of the BS.
The temporal distinguishability of photons in $a$ and $b$ is controlled by adjusting the time delay $\tau$ of the photons in $b$.
We observe two-photon interference (for single-photon Fock-states in $a$ and $b$) and four-photon interference (for two-photon Fock-states in $a$ and $b$) by two-fold and four-fold coincidence detection, respectively. See Methods for Fock-state postselection.
To investigate the questions posed above, we compare two different  detection schemes: either all four detectors are placed at mode $c$ only, which is used for the detection of (2,0) or (4,0) photons in modes $c$ and $d$ [\fig{fig:scheme}(a)], or three detectors are placed in  $c$ and one in $d$, allowing (1,1)- and (3,1)-detection [\fig{fig:scheme}(b)].

\begin{figure*}[t]
\centerline{\includegraphics[width=180mm]{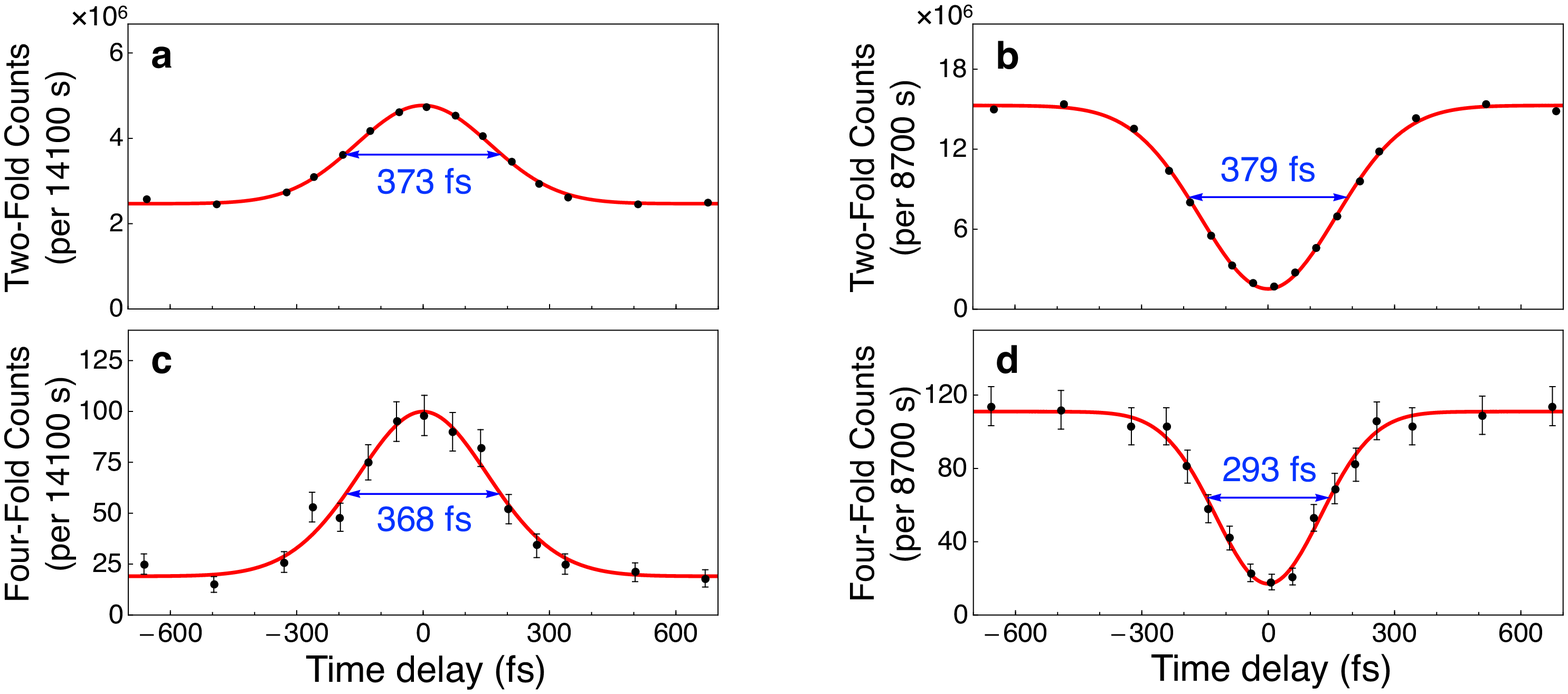}}
\caption{\textbf{Two- and four-photon interference observed under different detection schemes.} 
The detection schemes used are (2,0)-detection in \textbf{a}, (1,1)-detection \textbf{b}, (4,0)-detection \textbf{c}, and
(3,1)-detection \textbf{d}. Solid circles are experimental data, and solid lines are the best fits to the data using the fitting functions from our theoretical model, see \eq{eq:expprob}. Error bars represent one standard deviation from Poissonian counting statistics.
The (3,1)-detection signal exhibits a significant width reduction with respect to the (4,0)-detection signal.} \label{fig:data}
\end{figure*}

{\bf Experimental results.} First, two-photon interference of single-photon Fock-states 
 in input modes $a$ and $b$ is observed   \cite{PhysRevLett.59.2044,kim03a}. The detection of two photons at one output mode,  (2,0)-detection, is conducted by two-fold coincidence detection by D$_1$ and D$_3$ in the configuration shown in \fig{fig:scheme}(a). The coincidence counts exhibit a peak with a width of  $\sigma_{(2,0)}=373\pm7$ fs (we always refer to the full width at half maximum, FWHM), which is a manifestation of the two-photon coalescence effect, see \fig{fig:data}(a) \cite{kim03a}. The enhancement factor, i.e.~the ratio of the counts at $\tau=0$ and $\tau=\infty$, amounts to 1.93$\pm$0.02. On the other hand, single-photon detection at both output ports, i.e.~(1,1)-detection, is conducted by two-fold coincidence detection by  D$_1$ and D$_2$ in the configuration of \fig{fig:scheme}(b). We  observe a typical Hong-Ou-Mandel dip~\cite{shih2,PhysRevLett.59.2044}, shown in \fig{fig:data}(b), with a width of $ \sigma_{(1,1)}=379\pm5$ fs, which is very close to $\sigma_{(2,0)}$. As we shall show later in the analysis, this equality perfectly matches the theoretical expectation. 

A qualitatively different picture is encountered for four-photon interference. As described above, a two-photon Fock-state 
 is injected into each input mode ($a$ and $b$), and the four photons are detected under two different schemes: (4,0)-detection [\fig{fig:scheme}(a)] and (3,1)-detection [\fig{fig:scheme}(b)]. 
The (4,0)-signal presented in \fig{fig:data}(c) exhibits a peak with a width of ${\sigma_{(4,0)}=368\pm27 \text{ fs}}$ and an enhancement factor of 5.3$\pm$0.8, in agreement with the value 6 expected for four-photon bunching~\cite{PhysRevLett.83.959}. The width is comparable to the two-photon width, with $\sigma_{(4,0)}/\sigma_{(1,1)} = 0.97\pm0.07$.
The (3,1)-signal, on the other hand, exhibits a dip, as shown in \fig{fig:data}(d), since the four-photon path amplitudes that lead to (3,1)-events interfere destructively~\cite{PhysRevA.40.1371}. The width of the (3,1)-detection signal amounts to $\sigma_{(3,1)}= 293\pm13$ fs, the ratio of the widths is $\sigma_{(3,1)}/\sigma_{(1,1)}=0.77\pm0.04$. 
 Clearly, the dip in the (3,1)-signal is significantly narrower than the peak in the (4,0)-signal, while  the widths of the two-photon interference signals coincide, regardless of the detection scheme. We stress that these different widths result only from the choice of different detection events, since any other conditions such as spectral filtering are the same.

The width of multi-photon interference signals is usually interpreted as the coherence time of the interfering photons, but our above observations show that a four-photon state cannot be assigned a unique coherence time. In stark contrast to two-photon quantum interference, the choice of the observed detection event becomes relevant for the shape of the observed signal, as our subsequent analysis explains. 

{\bf Theoretical analysis.} For a quantitative understanding of the observed behavior, we operationally define the indistinguishability $\mathcal{I}(\tau)$ between two photons with time delay $\tau$ via the overlap of their amplitudes,
 \beq \mathcal{I}(\tau) =
\abs{\bra{0}\A(t+\tau)\A^\dagger(t)\ket{0}}^2, \label{eq:indistinguishability} \eeq
where $\A^\dagger(t)$ is a creation operator for a photon arriving at the BS centered at time $t$.
Two photons are fully indistinguishable for $\mathcal{I}=1$ and fully distinguishable for $\mathcal{I}=0$; for most quantum states, such as the frequently encountered Gaussian states (see Methods) these cases are attained for $\tau=0$ and $\tau=\infty$, respectively. 
When an ($N/2$)-photon Fock-state enters at each input mode of a BS, the state of the delayed photons in mode $b$ can be decomposed into a superposition of $N/2+1$ states, each term consisting of $d~(=0,1,\ldots, N/2)$ distinguishable and $N/2-d$ indistinguishable photons with respect to any photon in mode $a$ \cite{Ra:2011fk,PhysRevA.83.062111}. 
Each term that describes $d$ distinguishable photons at mode $b$, together with the ($N$/2)-photon Fock-state in mode $a$, is denominated of $d$-$distinguishability~type$. The weight of each type is a function of $\mathcal{I}(\tau)$ up to a power of $N/2$. 
As a result, the overall detection probability for the $(N-m,m)$ outcome is a polynomial in $\mathcal{I}(\tau)$ with a maximal power $N/2$,
\beq
P^{(N;N-m,m)}(\tau)&=&\sum_{k=0}^{N/2} c_k^{(N;N-m,m)}~\mathcal{I}(\tau)^k ,
\label{eq:expprob}
\eeq
as explained in more detail in the Methods section. 

\begin{table*}[t]
\begin{tabular}{lclll}
$N$~~&~~$(N-m,m)$~~~&$c_{k=0}~~~$&$c_{k=1}~~~$&$c_{k=2}~~~$\\
\hline \hline
2&$(2,0)$&$1/4$&$1/4$&$0$\\
&$(1,1)$&$1/2$&$-1/2$&$0$\\ \hline
4&$(4,0)$&$1/16$&$1/4$&$1/16$\\
&$(3,1)$&$1/4$&$0$&$-1/4$
\end{tabular}~~~~
\caption{
\textbf{Coefficient of k-th power of indistinguishability.} $c_k^{(N;N-m,m)}$ in \eq{eq:expprob} is given for the detection schemes shown in \fig{fig:scheme}.  For $k\geq3$, all coefficients vanish.
}\label{table:coeff}
\end{table*} 
With \eq{eq:expprob} at hand, we can understand the observed two-photon and four-photon interference signals: The different detection schemes in \fig{fig:scheme} result in distinct multi-particle interference paths \cite{PhysRevA.83.062111}, and, consequently, in distinct coefficients $c_k^{(N;N-m,m)}$, as shown in \tab{table:coeff}. As a consequence, the resulting signals differ.

In order to model the observations, we need to account for the Gaussian frequency distribution of the photons in our experimental setup.
 We denote the resulting width of the indistinguishability $\mathcal{I}(\tau)$ [\eq{eq:indistinguishability}] by  $\Delta \tau$, the width of higher orders $\mathcal{I}(\tau)^k$ is then reduced by a factor of $\sqrt{k}$ (see Methods).
 For two-photon interference, only the linear contribution $\mathcal{I}(\tau)$ contributes to the detection probabilities (\tab{table:coeff}); thus, both $P^{(2;2,0)}(\tau)$ and $P^{(2;1,1)}(\tau)$ exhibit the same width, and a unique coherence time can be assigned to the two-photon state. When four photons interfere, quadratic terms proportional to $\mathcal{I}(\tau)^2$ emerge in the detection probabilities. For $P^{(4;4,0)}(\tau)$, both $\mathcal{I}(\tau)$ and $\mathcal{I}(\tau)^2$ contribute, but the linear term has a larger pre-factor, such that it dominates the overall behavior. Consequently, the signal width is only slightly reduced with respect to two-photon interference, it amounts to approximately $0.93 \Delta \tau$,  in good agreement with the measured value  $\sigma_{(4,0)}/\sigma_{(1,1)} = 0.97\pm0.07$. On the other hand, for $P^{(4;3,1)}(\tau)$ only the quadratic term $\mathcal{I}(\tau)^2$ contributes and the resulting width amounts to $(1/\sqrt{2})\Delta \tau$, which is significantly narrower and coincides well with the experimentally observed value $\sigma_{(3,1)}/\sigma_{(1,1)}=0.77\pm0.04$. In other words, the second-order power of the indistinguishability $\mathcal{I}(\tau)$ determines the width of the multi-photon signal, and induces the hierarchy  $\sigma_{(3,1)} < \sigma_{(4,0)} < \sigma_{(1,1)} = \sigma_{(2,0)} = \Delta \tau$.

\begin{figure*}[t]
\centerline{\includegraphics[width=85mm]{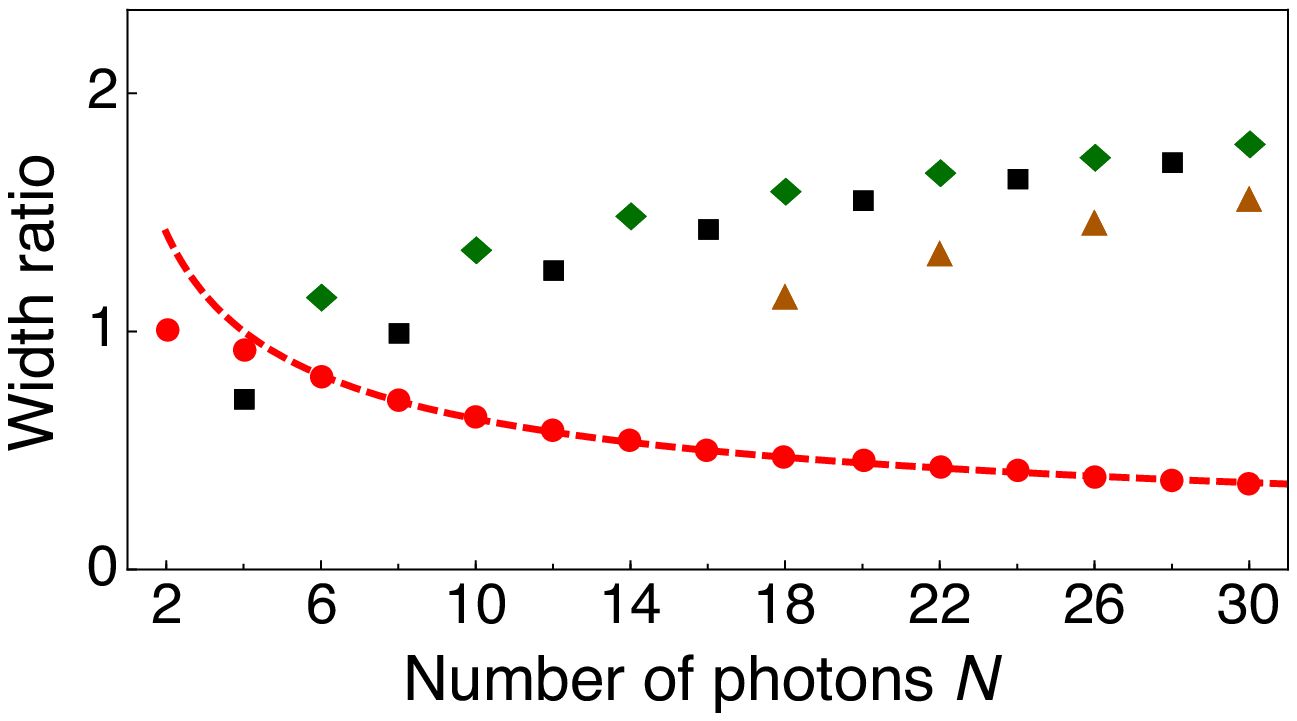}}
\caption{
\textbf{Relative multi-photon coherence times as a function of the number of photons.} Relative coherence times, quantified by width ratios $\sigma_{(N-m,m)}/\sigma_{(1,1)}$, are shown for photons with Gaussian frequency distribution [\eq{eq:gaussianfreq}]. Circles (red), diamonds (green), squares (black), and triangles (brown) represent the width ratios for the detection of $(N,0)$, $(N/2,N/2)$, $(N/2+1,N/2-1)$, and $(N/2+2,N/2-2)$, respectively. The dashed (red) line is the asymptotic expression $2/\sqrt{N}$ for the width ratio of the $(N,0)$-signal. }\label{fig:reduc}
\end{figure*}

\section{Discussion}
 When the number of interfering photons is increased further, higher orders of the indistinguishability become relevant. In $N$-photon interference, up to $N/2$-th order effects of the indistinguishability emerge in \eq{eq:expprob}, and each $k$-th power of the indistinguishability leads to a signal with a particular width. The coefficients $c_k^{(N;N-m,m)}$, which are determined by the chosen detection scheme \cite{PhysRevA.83.062111}, then regulate which power of the indistinguishability dominates the sum, and determine the observed multi-photon coherence time. Therefore, no unique coherence time can be assigned to a multi-photon state. The dependence on the detection scheme is an innate property of multi-photon states \cite{fortschritte2013}, which takes place irrespective of the specific shape of $\mathcal{I}(\tau)$.
 
   In particular, the multi-photon coherence time associated with  the detection of $(N,0)$ decreases with increasing number of photons $N$. For the Gaussian frequency distribution, the ratio $ \sigma_{(N,0)}/\sigma_{(1,1)}$  converges to the asymptotic value $2/\sqrt{N}$ (see Methods), shown as a red dashed line in Fig.~\ref{fig:reduc}.
On the other hand, coherence times related to the detection of the ${(N-m, m\neq0)}$-signal typically increase with the number of photons $N$. As an example, see the widths of the $(N/2,N/2)$-signal and of the $(N/2-2,N/2+2)$-signal for odd $N/2$, and the $(N/2+1,N/2-1)$-signal for even $N/2$ shown in Fig.~\ref{fig:reduc}. The increase of the signal width can be understood from the strong interference that is exhibited by  a Fock-state interfering with a single photon \cite{zyou2}: the $(N/2-1)$-distinguishability type 
 describes such a state with exactly one indistinguishable photon, which naturally features strong interference. With increasing particle number $N$, the ($N/2-1$)-distinguishability type becomes not only broader, but also more influential. 
 In general, the widths of $(N, 0)$ and $(N-m, m\neq0)$-signals exhibit opposite tendencies for increasing number of photons $N$, and the dependence of the observed multi-photon coherence times on the chosen event becomes more and more dramatic as more photons  interfere.

In conclusion, the possibility to measure the \emph{single}-photon coherence time via the width of the \emph{two}-photon interference signal turns out to be rather accidental: We have shown in experiment and in theory that the multi-photon coherence time, as defined by the width of the observed multi-photon signal, is not unique.
 Instead, it depends on the number of interfering particles and on the very way the photons are detected. 	
Thus, to a great extent, the strength of quantum multi-particle interference depends on the chosen observable.

\section{Methods}
{\bf Fock-states postselection.}
The quantum state of the photon pairs generated via SPDC
is proportional to $\sum_{j=0}^{\infty} \eta^j \ket{j,j}_{a,b}$ ,
 where $|\eta|^2$ is the single pair generation probability. Each component contains a Fock-state with an equal number of photons at the modes $a$ and $b$. Since $|\eta|^2=0.028$ in the experiment, two-fold and four-fold coincidences are attributed with high certainty to the respective states $\ket{1,1}_{a,b}$ and $\ket{2,2}_{a,b}$ at the input modes of the BS.

{\bf Wave-function decomposition.}
The quantum state of $N/2$ photons in mode $a$ and $N/2$ photons in the delayed mode $b$ is given as 
\beq 
\ket{\Psi}=  \frac{\left( \hat \A_{a}^\dagger(t) \right)^{N/2} }{\sqrt{(N/2)!}} \frac{\left( \hat \A_{b}^\dagger(t+\tau) \right)^{N/2}}{\sqrt{(N/2)!}} \ket{0} ,\label{eq:NphotonInitial}
\eeq
where $\hat \A_{a(b)}^\dagger(t)$ denotes the creation operator for a photon in mode $a(b)$. 
The creation operator for a delayed photon in $b$ can be written as a superposition of one indistinguishable (i.e.~parallel), and one distinguishable (i.e.~orthogonal) component with respect to a photon in $a$: 
\beq 
\hat \A_{b}^\dagger(t + \tau) =\sqrt{\mathcal{I}(\tau)}~\hat \A_{b}^\dagger(t)+ \sqrt{1-\mathcal{I}(\tau)}~\hat \B_b^\dagger(t,\tau) ,\label{eq:orthonomalization}
\eeq 
where $\mathcal{I}(\tau)$ is the indistinguishability given in \eq{eq:indistinguishability} and $\hat \B_b^\dagger(t,\tau)$ is the creation operator that takes into account the part of the wave-function in $b$ that is orthogonal to the wave-function of a photon in $a$. In other words, each photon in $b$ is indistinguishable with respect to the photons in $a$ with probability $\mathcal{I}(\tau)$. By substitution of \eq{eq:orthonomalization}, the multi-photon state in \eq{eq:NphotonInitial} becomes 
\beq 
\ket{\Psi}=  \frac{1}{(N/2)!} \left( \hat \A_{a}^\dagger(t) \right)^{N/2}   \sum_{d=0}^{N/2}  { N/2 \choose d}  \left( \sqrt{\mathcal{I}(\tau)} ~\hat \A_{b}^\dagger(t) \right)^{N/2-d}  \left( \sqrt{1-\mathcal{I}(\tau)}~ \hat \B_b^\dagger(t,\tau) \right)^d  ,
\label{eq:NphotonDelayed}
\eeq
where we can identify $N/2+1$ different \emph{d-distinguishability types}, i.e.~terms with a fixed number $d$ of distinguishable and $N/2-d$ of indistinguishable photons in $b$ with respect to the photons in $a$, together with the $N/2$ photons in $a$~\cite{Ra:2011fk,PhysRevA.83.062111}. For each type, $N/2-d$ photons in $b$ interfere with the photons in $a$, while the remaining $d$ photons do not.
The weight of the $d$-distinguishability type is a polynomial in $\mathcal{I}(\tau)$, given as
\beq
 W_d^{(N)}(\tau)= {N/2 \choose d}~\mathcal{I}(\tau)^{N/2-d} \left(1-\mathcal{I}(\tau)\right)^{d},
\label{eq:weight}
\eeq
which can be deduced from \eq{eq:NphotonDelayed}.

{\bf Overall detection probability.}
Since each $d$-distinguishability type is orthogonal to all other distinguishability types, the overall detection probability for the  outcome $(N-m,m)$ is the mean of the detection probabilities associated with each $d$-distinguishability type $p_{d}^{(N;N-m,m)}$~\cite{PhysRevA.40.1371, ou06, zyou}, weighted by $W_d^{(N)}(\tau)$:
\beq
P^{(N;N-m,m)}(\tau)&=&\sum_{d=0}^{N/2} p_{d}^{(N;N-m,m)}~
W_{d}^{(N)}(\tau).
\label{eq:detprob}
\eeq
Since $W_d^{(N)}(\tau)$ contains $\mathcal{I}(\tau)$ up to a power of $N/2$, the overall detection probability is a polynomial in $\mathcal{I}(\tau)$ with a maximal power of $N/2$, which can be expressed by \eq{eq:expprob}.

In particular, the probability for $(N,0)$-detection results in a $k$-th order coefficient in \eq{eq:expprob} corresponding to 
the square of a binomial coefficient,
\beq c_k^{(N;N,0)}=2^{-N} {N/2 \choose k}^2 , \eeq
such that the most important contribution is given by $\mathcal{I}(\tau)^{k=N/4}$, which quickly becomes dominant  as $N$ increases.
 As a result, the width of the $(N,0)$-signal converges to the width of the $N/4$-th order of the indistinguishability, in the limit of large $N$.

{\bf Photons with Gaussian frequency distribution.}
A photonic creation operator with Gaussian frequency distribution is given by
\begin{align}
\A^\dagger(t)=\frac{1}{\sqrt \pi\Delta \omega} \int \mbox{d} \omega \exp\left(- \frac{(\omega-\omega_0)^2}{2 \Delta\omega^2}+i \omega t\right)  \A_{\omega}^\dagger \ , \label{eq:gaussianfreq}
\end{align}
where $t$ is the central arrival time of the photon at the BS and $\A_{\omega}^\dagger$ is the creation operator for a photon of frequency $\omega$; the central frequency is $\omega_0$ and the bandwidth $\Delta\omega$. 
In this case, the indistinguishability [\eq{eq:indistinguishability}] becomes
$\mathcal{I}(\tau) =
\exp\left({-\Delta\omega^2 \tau^2 /2}\right)$, 
it is characterized by a FWHM of  $\Delta \tau =\sqrt{8 \ln 2}/\Delta\omega$.
The $k$-th power of the indistinguishability, $\mathcal{I}(\tau)^k =
\exp\left({-k\Delta\omega^2 \tau^2 /2}\right)$, 
 has a reduced FWHM of $(1/\sqrt{k}) \Delta\tau$.

\clearpage


\clearpage

\section{Acknowledgements}
 This work was supported in part by the National Research Foundation of Korea (grant no. 2013R1A2A1A01006029 and 2011-0021452) and the DAAD/GEnKo Grant 50739824. M.C.T. acknowledges the financial support from the German National Academic Foundation, by the Alexander von Humboldt Foundation through a Feodor Lynen Fellowship, and by the bilateral DAAD-NRF scientist exchange programme. H.T.L. acknowledges the financial support from the National Junior Research Fellowship (2012-000642). A.B. acknowledges partial support through the EU-COST Action MP1006 ``Fundamental Problems in Quantum Physics" and DFG Research Unit 760.

\section{Author contributions}
Y.-S.R., M.C.T., F.M., A.B., and Y.-H.K. conceived the idea; Y.-S.R., H.-T.L., and O.K. performed the experiment; 
Y.-S.R., H.-T.L., and Y.-H.K. analyzed the data. All authors discussed the results and contributed to writing the manuscript.

\newpage


\begin{thebibliography}{10}
\newcommand{\enquote}[1]{``#1''}

\bibitem{PhysRevLett.59.2044}
Hong, C. K.,  Ou, Z. Y. \& Mandel, L. Measurement of subpicosecond time intervals between two photons by interference. \textit{Phys. Rev. Lett.} \textbf{59}, 2044-2046 (1987).

\bibitem{kim03a} Kim, Y.-H. \& Grice, W. P. Observation of correlated-photon statistics using a single detector. \textit{Phys. Rev. A} \textbf{67}, 065802 (2003).

\bibitem{Bocquillon} Bocquillon, E. \textit{et al.} Coherence and indistinguishability of single electrons emitted by independent sources. \textit{Science} \textbf{339}, 1054-1057 (2013).

\bibitem{Hornberger2012} Hornberger, K., Gerlich, S., Haslinger, P., Nimmrichter, S. \& Arndt, M. Colloquium: quantum interference of clusters and molecules. \textit{Rev. Mod. Phys.} \textbf{84}, 157-173 (2012).

\bibitem{keller97} Keller, T. \& Rubin, M. H. Theory of two-photon entanglement for spontaneous parametric down-conversion driven by a narrow pump pulse. \textit{Phys. Rev. A} \textbf{56}, 1534-1541 (1997).

\bibitem{shih2} Shih, Y. H. \& Alley, C. O. New type of Einstein-Podolsky-Rosen-Bohm experiment using pairs of light quanta produced by optical parametric down conversion. \textit{Phys. Rev. Lett.} \textbf{61}, 2921-2924 (1988).

\bibitem{ref:pittman} Pittman, T. B. \textit{et al.} Can two-photon interference be considered the interference of two photons? \textit{Phys. Rev. Lett.} \textbf{77}, 1917-1920 (1996).

\bibitem{kim03} Kim, Y.-H. Two-photon interference without bunching two photons. \textit{Phys. Lett. A} \textbf{315}, 352-357 (2003).

\bibitem{kwiat92} Kwiat, P. G., Steinberg, A. M. \& Chiao, R. Y. Observation of a ``quantum eraser'': a revival of coherence in a two-photon interference experiment. \textit{Phys. Rev. A} \textbf{45}, 7729-7739 (1992).

\bibitem{PhysRevLett.68.2421} Steinberg, A. M., Kwiat, P. G. \& Chiao, R. Y. Dispersion cancellation in a measurement of the single-photon propagation velocity in glass. \textit{Phys.  Rev. Lett.} \textbf{68}, 2421-2424 (1992).

\bibitem{PhysRevLett.83.959}
Ou, Z. Y., Rhee, J.-K. \& Wang, L. J. Observation of four-photon interference with a beam splitter by pulsed parametric down-conversion. \textit{Phys. Rev. Lett.} \textbf{83}, 959-962 (1999).

\bibitem{Niu:09}
Niu, X.-L. \textit{et al.} Observation of a generalized bunching effect of six photons. \textit{Opt. Lett.} \textbf{34}, 1297-1299 (2009).

\bibitem{PhysRevA.40.1371}
Campos, R. A., Saleh, B. E. A. \& Teich, M. C. Quantum-mechanical lossless beam splitter: SU(2) symmetry and photon statistics. \textit{Phys. Rev. A} \textbf{40}, 1371-1384 (1989).

\bibitem{Ra:2011fk}
Ra, Y.-S. \textit{et al.} Nonmonotonic quantum-to-classical transition in multiparticle interference. \textit{Proc. Natl. Acad. Sci.  USA} {\bf 110}, 1227-1231 (2013).

\bibitem{PhysRevA.83.062111}
Tichy, M. C. \textit{et al.} Four-photon indistinguishability transition. \textit{Phys. Rev. A}  \textbf{83}, 062111 (2011).

\bibitem{fortschritte2013}
Tichy, M. C., De Melo, F., Ku\'s, M., Mintert, F. \& Buchleitner, A., Entanglement of identical particles and the detection process. \textit{Fortschr. Phys.} {\bf 61}, 225-237 (2013).

\bibitem{zyou2}
 Ou, Z. Y. Quantum multi-particle interference due to a single-photon state. \textit{Quantum Semiclass. Opt.} \textbf{8}, 315-322 (1996).

\bibitem{ou06}
Ou, Z. Y. Temporal distinguishability of an N-photon state and its characterization by quantum interference. \textit{Phys. Rev. A} \textbf{74}, 063808 (2006).

\bibitem{zyou}
Ou, Z. Y. \textit{Multi-Photon Quantum Interference}, 185-224 (Springer, New York, 2007).

\end{thebibliography}
\end{document}